\documentclass{article}
\usepackage{icmc2021template}
\usepackage{times}
\pdfoutput=1
\usepackage{ifpdf}
\usepackage{soul}
\usepackage{multirow}
\usepackage{subcaption}
\usepackage{url} 
\usepackage{cite}
\usepackage[numbers, sort, comma, square]{natbib}

\def\papertitle{Performer Identification From Symbolic Representation of Music Using Statistical Models}
\def\firstauthor{Syed Rifat Mahmud Rafee}
\def\secondauthor{Gy{\"o}rgy Fazekas}
\def\thirdauthor{Geraint A.~Wiggins}

\newif\ifpdf
\ifx\pdfoutput\relax
\else
   \ifcase\pdfoutput
      \pdffalse
   \else
      \pdftrue
  \fi
\fi

\ifpdf 
  \usepackage[pdftex,
    pdftitle={\papertitle},
    pdfauthor={\firstauthor, \secondauthor, \thirdauthor},
    bookmarksnumbered, 
    pdfstartview=XYZ 
   ]{hyperref}

  \usepackage[pdftex]{graphicx}
  \graphicspath{{./figures/}}
  \DeclareGraphicsExtensions{.pdf,.jpeg,.png}

  \usepackage[figure,table]{hypcap}

\else 
  \usepackage[dvips,
    bookmarksnumbered, 
    pdfstartview=XYZ 
  ]{hyperref}  

  \usepackage[dvips]{epsfig,graphicx}
  \graphicspath{{./figures/}}
  \DeclareGraphicsExtensions{.eps}

  \usepackage[figure,table]{hypcap}
\fi

\hypersetup{
    colorlinks,%
    citecolor=black,%
    filecolor=black,%
    linkcolor=black,%
    urlcolor=black
}

\title{\papertitle}

 \threeauthors
  {0.5in}
  {\firstauthor} {Queen Mary University of London\hspace*{-.3cm}\\~\\ %
     {\tt \href{mailto:s.rafee@qmul.ac.uk}{s.rafee@qmul.ac.uk}}}
  {\secondauthor} {Queen Mary University of London \\~\\ %
     {\tt \href{mailto:g.fazekas@qmul.ac.uk}{g.fazekas@qmul.ac.uk}}}
  {\thirdauthor} { Vrije Universiteit Brussel \&\\\hspace*{-.3cm}Queen Mary University of London\\ %
     {\tt \href{mailto:geraint@ai.vub.ac.be}{geraint@ai.vub.ac.be}}}

\begin{document}
\capstartfalse
\maketitle
\capstarttrue

\begin{abstract}
 Music Performers have their own idiosyncratic way of interpreting a musical piece. A group of skilled performers playing the same piece of music would likely to inject their unique artistic styles in their performances. The variations of the tempo, timing, dynamics, articulation etc. from the actual notated music are what make the performers unique in their performances. This study presents a dataset consisting of four movements of Schubert's ``Sonata in B-flat major, D.960" performed by nine virtuoso pianists individually. We proposed and extracted a set of expressive features that are able to capture the characteristics of an individual performer's style. We then present a performer identification method based on the similarity of feature distribution, given a set of piano performances. The identification is done considering each feature individually as well as a fusion of the features. Results show that the proposed method achieved a precision of 0.903 using fusion features. Moreover, the onset time deviation feature shows promising result when considered individually.
\end{abstract}

\vspace{-5pt}
\section{Introduction}\label{sec:introduction}
\sloppy
Music performance is considered to be a creative process. Musicians often express their emotions to the audience through music. The perceived emotions are subjective to the audiences as the interpretations of the observed music may vary from person to person \cite{barthet2012music,russel}. Virtuoso performers are often recognized by people for their uniqueness in portraying their emotional expressions in a music performance. In music, these unique emotional expressions are characterised by some parameters such as tempo, dynamics and articulation. And variations in these parameters make the performers distinguishable and unique in their performances. 

Expressive performance has been a central research topic in contemporary musicology, hence a large research body has been devoted to exploring expressive piano performance, by examining the performance parameters that are used by pianists. So far, expressive timing and amplitude have been the most explored features for their salience and effect on the perception of emotional expression \cite{perceptionofemotionalexpression}. \cite{STAMATATOS200537} developed a dataset consisting of 22 performers and used expressive features like timing, loudness and articulation extracted from MIDI performances, for automatic performer identification using machine learning models. They have introduced `norm deviation' features along with the `score deviation' features which shows that it is indeed possible for a machine to distinguish performers based on their performance style. The Authors in \cite{wang,saunder} used the same features but extracted from audio recordings of piano performances. \cite{Rink2011MotiveGA} has studied individuality in expressive piano performance in light of musical gestures, i.e., timing and dynamic patterns which can characterize the individual expressive strategies. \cite{pianistindi} studied different timbral features to investigate individuality in piano performances. They studied the pattern of timbral expressions induced by four pianists. The result shows that pianists can express individual style through specific timbral intentions. 

In this paper, we study several statistical models and test them in the performer identification task by quantifying expressive performance parameters that are
important in expressing the individuality of piano performers \cite{STAMATATOS200537,Rink2011MotiveGA,pianistindi}. In other words, we use statistical models to identify piano performers given a set of performances of the same piece of music by several virtuoso pianists. This can be viewed as a classification problem where the classes are the candidate pianists. We compare different classification methods using features that are extracted from `norm performance' introduced in \cite{STAMATATOS200537}. Here, `norm performance' refers to the average of the onset time, off time and dynamic of each note across all performances in the dataset. This norm performance is then considered as a reference point. The main performance features that are accounted in this study are timing (variations in tempo), dynamic level (variations in loudness) and articulation (the use of overlaps and pauses). We conducted classification considering each feature individually as well as a feature fusion method to consider the features in combination. We also propose note duration feature which demonstrates promising result when fused with other features. Our experimental result shows that it is possible to distinguish pianists using the proposed features despite the limited training data.

The rest of the paper is organised as follows: Section \ref{dataset} introduces a novel dataset consisting of 16980 notes for each performer. Section \ref{methodology} describes the feature data-prepossessing which includes MIDI-to-MIDI alignment, feature extraction and selection process. In Section \ref{experiments}, we discussed the classification experiments and results and Section \ref{discussion} outlines the conclusion and possible future developments.

\vspace{-5pt}
\section{DATASET} \label{dataset}

The data used in this study consists of performances played and recorded on a Yamaha CFX concert Grand Piano which is equipped with state-of-the-art Disklavier Pro recording technology. We chose 9 virtuoso pianists from the International Piano-e-competition \cite{epiano} who played the same four movements of Sonata in B-flat Major, D960, I. Molto moderato, II. Andante sostenuto, III. Scherzo: Allegro vivace con delicatezza – Trio and IV. Allegro ma non troppo – Presto composed by Franz Schubert (see Table \ref{tab:datasets}). This is one of the last major compositions by Schubert for solo piano and considered among the most important of Schubert's mature masterpieces. Each movement has a different number of notes as described in Table \ref{tab:datasets}. 
\begin{table}[]
\resizebox{\columnwidth}{!}{
\begin{tabular}{|c|c|c|c|}
\hline
\textbf{Composer}   &\textbf{Piece} &\textbf{Movement}  &\textbf{No.of Notes} \\ 
\hline
\multirow{4}{*}{F.Schubert} & \multirow{4}{*}{\shortstack{Sonata in\\ B-Flat Major,D960}}  &   I   &   7582 \\
\cline{3-4}
                            &   &   II  &   2005 \\
                            \cline{3-4} 
                            &   &   III &   2717 \\
                            \cline{3-4} 
                            &   &   IV  &   4676 \\
                            \hline
\end{tabular}
}
\caption{Details of Dataset with number of notes.}\label{tab:datasets}
\end{table}

All the performances have been recorded in both raw audio and MIDI format. In this experiment, we are mostly interested to work with the symbolic format of the music which is MIDI. The advantage of MIDI over raw audio is that raw audio needs manual annotation of the note events. This is very time consuming and requires a lot of expertise. However, in MIDI, each note event is explicitly described by four parameters: Onset (note on time), Offset (note-off time), Pitch (numerical value ranges from 0 to 128) and dynamic level (loudness of the note, ranges from 0 to 128). Hence, manual annotation of notes is no longer required.

\vspace{-5pt}
\section{METHODOLOGY} \label{methodology}
In order to explore pianist's individuality in expressive performances, our study is designed with respect to the following steps: Alignment of performed musical pieces based on their notes to calculate the `norm performance', extraction of meaningful norm based performance parameters and finally the classification process using the performance features. 

\begin{figure}[b]
\centering{\includegraphics[width=7cm]{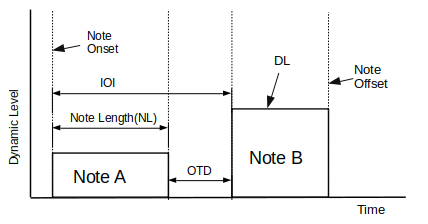}}
\caption{\label{fig:feat}Five parameters that are considered to characterize note-level performance details}
\end{figure}

\vspace{-5pt}
\subsection{MIDI-to-MIDI Alignment}
Symbolic music alignment is a process of automatically matching a note in a music performance with a corresponding note in a score or a reference performance. We use a HMM (Hidden Markov Model) based symbolic music alignment algorithm proposed by \cite{nakamura2017performance}. The algorithm achieved high accuracy with a low computational cost. Initially it aligns two signals, reference and performance, using hidden Markov models (HMMs) and detects the performance errors from initial alignment result and then uses a merged-output HMMs \cite{nakamura} to correct the errors as a post-processing step.


There are two modes of the alignment process: the first one is, Score-to-MIDI alignment where the score file is a MusicXML file with no performance parameters and the second one is the MIDI-to-MIDI alignment. The MIDI-to-MIDI alignment algorithm uses any two midi files to find the corresponding notes between them. One of them can be used as a reference signal and the other can be used as the performance signal. The reference signal is first converted into a score file and the score-to-MIDI alignment algorithm is then used for the converted score and the performance MIDI file. 
\begin{figure}[]
 \centering
 \includegraphics[width=7cm,height=3.2cm]{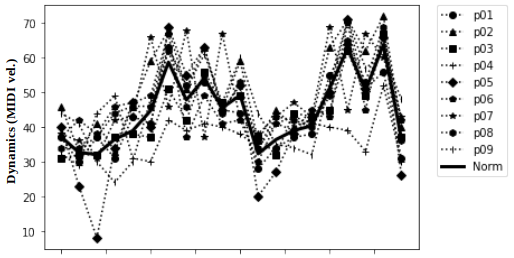}
 \caption{\label{fig:norm}Dynamic variations for the first 20 notes of the Sonata in B flat major,D960, Mvmnt. II, as performed by pianists p01-p09}
\end{figure}

\vspace{-5pt}
\subsection{Feature Extraction}
\begin{figure}[]
 \centering
 \includegraphics[width=7cm,height=3.2cm]{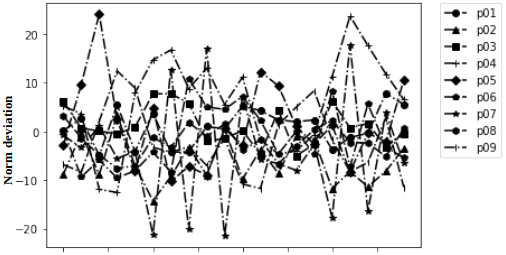}
 \caption{\label{fig:norm_dv} Velocity deviations of first 20 notes from norm performance as performed by pianists p01-p09.}
\end{figure}

Expressive performance can be defined as, either the intended deviation from the score, which is purely a mechanical rendition of the musical piece in terms of tempo, dynamic and articulation or the average performance considering it as a reference point. Previously, \cite{Stamatatos2001ACM} have used score deviation features to successfully discriminate two skilled performers playing the same piano piece. But later a comprehensive and empirical study in \cite{stamatatos2} revealed that deviation from the average performance is more powerful in representing performer's individuality than the deviation from the printed score. Also, the empirical results shows that the norm based features are proved to be very accurate for intra-piece tests (training and test set taken from the same piece) and inter-piece tests (training and test set taken from different pieces). We take the later approach which is to calculate the note level performance deviations from an average performance. The idea of performance norm can be better understood from Figure \ref{fig:norm}, where the bold line denotes the average performance calculated from pianists p01-p09 in terms of note dynamics. We can see that, the norm performance follows a basic form of individual performances but as illustrated in Figure \ref{fig:norm_dv} the deviations from the norm does not necessarily follow a similar form (peaks and dips of notes) and the differences are clearly highlighted.

\vspace{-5pt}
\subsubsection{Proposed Features} \label{proposed_features}

In order to represent the stylistic properties of a performer from their performance, we propose the following set of features, given the performance norm derived from the set of different performers: 

 \begin{itemize}
    \item[$-$] Norm deviation features:
    \begin{itemize}
        \item[$\bullet$] d(OT$_{n}$,OT$_{p}$) 
        \item[$\bullet$] d(IOI$_{n}$,IOI$_{p}$) 
        \item[$\bullet$] d(OTD$_{n}$,OTD$_{p}$) 
        \item[$\bullet$] d(DL$_{n}$,DL$_{p}$) 
        \item[$\bullet$] d(ND$_{n}$,ND$_{p}$) 
    \end{itemize}    
\end{itemize}
 Here, d(x$_{n}$,y$_{p}$) denotes the deviation of a vector of numerical values y from a vector of values x where n and p refers to the reference vector and the actual performance respectively. The parameters can be better understood from Figure \ref{fig:feat}, where, OT (Onset time) denotes the start of a note, IOI (Inter Onset Interval) denotes the time interval between the onsets of two notes, OTD (Offset Time Duration) denotes the time interval between the end of a note and the onset of the next note, DL (Dynamic level) denotes the loudness of a note and ND (Note duration) denotes the exact duration of a note. In addition, we calculated the cross-correlation between the individual features using Pearsons correlation coefficient (r). We found that OTD and ND to be highly correlated (r$>$0.9) with each other. However, we used both of the features in our study to find the most effective feature for performer identification task.
 
 There are other performance related features that could be considered but ignored in this study. This includes melody lead \cite{melodylead} which is used by music performers to emphasize a voice over others. An empirical study was conducted by \cite{mlead} which shows that melody lead has a great impact on expressiveness. This constitutes future work for us. Different distance types can be considered in order to measure the deviation in each features defined in the previous subsection. The distance metrics that are considered in this study are formulated below:

\begin{equation} \label{eq1}
Simple: d_s(x,y) = (x_i - y_i)
\end{equation}

\begin{equation} \label{eq2}
Simple Absolute: d_{ab}(x,y) = (|x_i| - |y_i|)
\end{equation}

The experimental result shows that, the simple distance metric formulated in equation \ref{eq1} best fits for the OT deviation, DL deviation and ND deviation features where the simple absolute equation \ref{eq2} best fits for the IOI and OTD deviation features.
\subsection{Pianist Classification}
In this section, we describe the pianist classification that includes the feature distribution estimation and similarity calculation. The identification method is inspired by the method introduced in \cite{masterviolin}. We first discuss the methods in terms of each features and then, we discuss a feature fusion technique in Section \ref{feature_fusion_method}.

\vspace{-5pt}
\subsubsection{Using Features Individually}
As depicted in Figure \ref{fig:norm}, the dynamic variation for 9 performers who played the same piece of music shows that different performers playing the same music will always have their own way of expressing the piece to the audience. The average performance line in bold indicates the performers understanding of the structure of the music while the individual peaks and dips shows how much each performer deviates from the average performance. These deviations characterise each performer's individually. Therefore, we model these distinct characteristics of performers using the distribution of each features proposed in subsection \ref{proposed_features}. To model these deviation distributions we use Histogram, Kernel densities (KDE) and Gaussian Mixture Models (GMMs) similarly to \cite{masterviolin, violinist}. We assume that, the distribution of the deviations provide us compact representation of performers idiosyncratic style. 

\begin{figure}[]
\centering
  \begin{subfigure}[b]{3.6cm}
    \includegraphics[width=3.6cm]{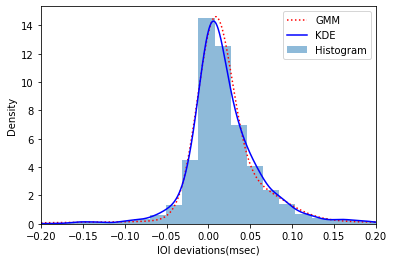}
    \caption{Pianist 1}
    \label{fig:ioi_p1}
  \end{subfigure}
  \begin{subfigure}[b]{3.6cm}
    \includegraphics[width=3.6cm]{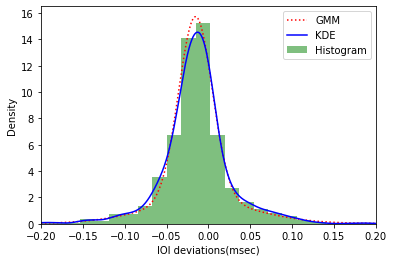}
    \caption{Pianist 2}
    \label{fig:ioi_p2}
  \end{subfigure}
  \caption{\label{fig:ioi_dev}Distribution of two performer's Inter-Onset Interval (IOI) deviations.}
\end{figure}

Expressive timing is one of the most important performance features for performer identification. It has great impact on performers playing style. Performers vary the tempo throughout the performance to express their emotions distinctively. Figure \ref{fig:ioi_dev} shows the inter-onset interval deviation distribution of two performers calculated using Histogram, KDE and GMMs. We can observe that deviations are mostly concentrated between -0.01 ms and 0.03 ms for pianist 1 whereas it is between -0.04 ms and 0 ms for pianist 2. This indicates that pianist 1 prefers to perform with faster tempo whereas pianist 2 prefers to perform with relatively slower tempo than player 1. Based on the similar observations with the rest of the features, onset time deviation, off-time duration deviation, dynamic level deviation and note duration deviation, it can be presumed that these features can reflect important aspects of the pianist's individual characteristics. The red dotted line in Figure \ref{fig:ioi_dev} represents the Gaussian Mixture Models(GMMs). We trained the GMMs with 2,3,5 and 7 components and the empirical results show that GMMs do not require more than 3 components to model the distributions. The red dotted PDF curve of the GMMs in Figure \ref{fig:ioi_dev} also shows that, it represents the Histograms and KDEs well. The blue line in Figure \ref{fig:ioi_p1} shows the PDF of the KDE. The bandwidths of KDEs to model the OT deviation, IOI deviation, OTD deviation, DL deviation and ND deviations are 1.2, 0.01, 0.02, 1.5 and 0.01 respectively. This Figure also shows that, the PDF curves of both performers show similar properties to the histogram. 

In order to quantify the differences between performers from the distributions, we take a similarity measurement step of each feature distribution for every performer in our dataset using Kullback-Leibler (KL) divergence\cite{kullback1951} as shown below:
\begin{equation} \label{eq3}
D_{KL}(P{\parallel}Q) = \sum_{x\in \chi}P(x)log(\frac{P(x)}{Q(x)})
\end{equation}
The KL-divergence which is also known as relative entropy, is a measure of how one probability distribution is different from another probability distribution or in other words, from equation \ref{eq3} we can say, it computes the likelihood ratio between two distributions and tells how probability distribution Q diverges from the probability distribution P by computing the cross-entropy minus the entropy. We measure the KL-divergence for Kernel densities using the approach introduced in \cite{sasaki2014direct}. We also calculate the KL-divergence of two GMMs, but since the KL-divergence for two GMMs has no closed form expression, it is not analytically tractable. Hence, we use a variational Bayes approximation method \cite{4218101} to avoid this issue.

We use KL-divergence to calculate the divergence between an unknown performer's
feature distribution and every known performer's feature distribution in the dataset in order to classify the unknown performer. The minimum divergence between the unknown and the known performer's distribution classifies the unknown performer.

\subsubsection{Feature Fusion Method} \label{feature_fusion_method}
Besides classification using individual features, we use feature fusion techniques that are able to combine multiple features (see subsection \ref{proposed_features}). Not all features will have a similar importance for any classification task. Features with less importance may lead to a suboptimal classification result. Hence, weights can be used to increase the influence of good features and decrease the influence of bad features. In this study, we used linear combination with equal weights to fuse similarity estimates for the distributions of different features. Optimizing weights as done in \cite{fusion,optweights1,optweights2,Sheng2018FeatureSF}, for individual features could be a potential future work.
We use Leave one group out cross validation technique with 8 folds to calculate the KL-divergence between the training and testing set for every group of data. We then combine the similarity estimation of feature distributions in each fold for different types of features. This is formulated in equation \ref{eq4}:
\begin{equation} \label{eq4}
KL_{total} = \sum_{i=1}^{|N|}w_{i}KL_{N_i}
\end{equation}
where, N = \{N1, N2, N3, N4, N5\} denotes the set of statistical models corresponds to OT deviation, IOI deviation, OTD deviation, DL deviation and ND deviation features respectively (see subsection \ref{proposed_features}) which are computed separately. \(w_i\) denotes the corresponding feature weight which is set to 1 in our experiment. However, the feature fusion technique used in this study is not unique. We can combine any 2, 3, 4 or 5 features together to calculate the overall KL-divergence. In the next section, we describe several experiments with both single feature and fused features to validate the pianist identification methods and determine how accurate the methods are for pianist classification.
\begin{table}[h]
\centering

\resizebox{\columnwidth}{!}{%
\begin{tabular}{|l|c|c|c|c|c|}
\hline
\multicolumn{1}{|c|}{\multirow{2}{*}{Model}} & \multicolumn{5}{c|}{Feature}                                                                                                       \\ \cline{2-6} 
\multicolumn{1}{|c|}{}                       & \multicolumn{1}{l|}{OT} & \multicolumn{1}{l|}{IOI} & \multicolumn{1}{l|}{OTD} & \multicolumn{1}{l|}{DL} & \multicolumn{1}{l|}{ND} \\ \hline
Histogram                                    & 0.603                        & 0.459                        & \textbf{0.457}                        & 0.533                       & 0.398                       \\ \hline
KDE                                          & \textbf{0.688}                       & \textbf{0.489}                       & 0.342                        & \textbf{0.589}                       & \textbf{0.498}                       \\ \hline
GMM                                          & 0.591                        & 0.476                        & 0.267                        & 0.564                       & 0.405                       \\ \hline
\end{tabular}%
}
\caption{Pianist identification result in terms of Precision.}
\label{tab:table2}
\end{table}

\vspace{-5pt}
\section{EXPERIMENTS} \label{experiments}
In this section, we show the performer classification results based on each feature: note onset time deviation, inter-onset interval deviation, off-time duration deviation, dynamic level deviation and note duration deviation as well as the combination of these features using the fusion method described in Section \ref{feature_fusion_method}. We also verify the effectiveness of the proposed performer identification methods using leave one group out cross validation technique. We then calculate the F-measure which is a way to combine both precision and recall into single measure that captures both the properties. A normalized confusion matrix is also used to show the performance of identification for all the performers in our dataset.

\begin{table}[]
\centering

\scalebox{0.9}{
\begin{tabular}{|c|c|c|c|}
\hline
Feature & Precision & Recall & F-score \\ \hline
OT               & \textbf{0.688}     & \textbf{0.667}  & \textbf{0.677}   \\ \hline
IOI              & 0.489              & 0.472           & 0.480            \\ \hline
OTD              & 0.342              & 0.375           & 0.357            \\ \hline
DL               & 0.589              & 0.542           & 0.564            \\ \hline
ND               & 0.498              & 0.458           & 0.477            \\ \hline
\end{tabular}%
}
\caption{Pianist identification result using KDE. Numbers in bold denote the best result.}
\label{tab:table3}
\end{table}

\subsection{Classification result based on individual features}

In our experiment, a total of 16980 aligned notes has been extracted from each pianist's performance. Since, we consider the note level local deviation features in our experiment, the same amount of deviation features are also extracted from the notes. We perform leave one group out cross validation to separate each performers data into 8 groups to maintain a high number of cross validation folds as well as to ensure there are enough data in every test set.

Hence, 2122 notes are designated for each of the first 7 folds and the last fold contains 2126 notes. We then select a random performer out of the 9 performers, and designate one group of data from that performer as test data. The rest of the groups of all 
\begin{figure}[h]
\centering
\includegraphics[width=7cm]{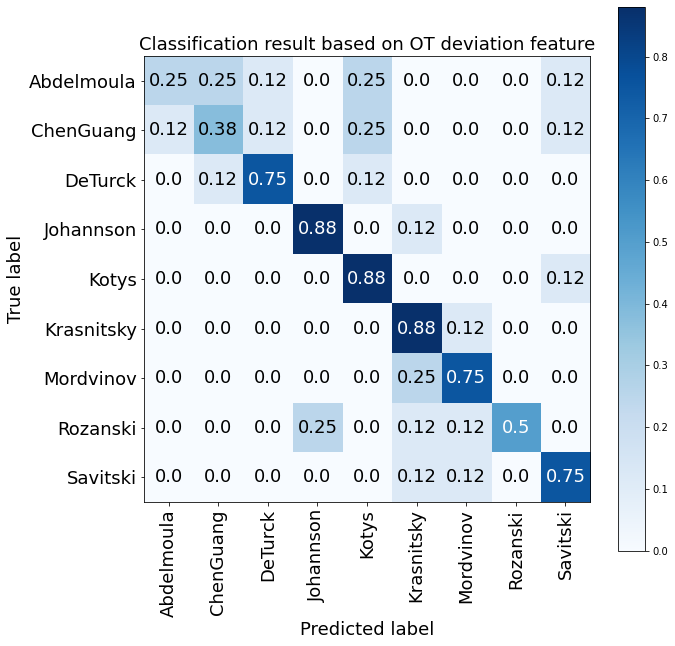}
\caption{\label{fig:ot_dev_class}Normalized confusion matrix for pianist classification using OT deviation.}
\end{figure}
performers are considered as training data. The distributions of both the test and training set are calculated using Histogram, KDE and GMMs. The Histogram bins, Kernel density bandwidth and the GMM hyper-parameters are optimised in this experiment and optimum values for each feature are kept constant.

Finally, we calculate the KL-divergence between the test distribution and the training distribution for every performer in the dataset in order to measure the similarity. The minimum KL-divergence value identifies the unknown performer. In other words, we can say that, the corresponding performer's training distribution which has the minimum distance with the test distribution is the identified performer. Table \ref{tab:table2} shows the identification result in terms of precision for each feature. The results in bold number denotes the best precision achieved by the models for each feature. We can see that the OT deviation feature performs best when considered individually out of all the features for all the models. And, KDE performs the best out the other two distribution models. A normalized confusion matrix is also shown in Figure \ref{fig:ot_dev_class} for OT deviation using KDE. The x-axis corresponds to the predicted performers label and the y-axis corresponds to the true performers label.

\subsection{Classification Result Using the Fusion of Features}
As discussed in Section \ref{feature_fusion_method}, feature fusion is a method of combining two or more features to remove redundant and irrelevant features for better classification accuracy. In this section we combine two or more features and fuse them together to make a new feature set which are tested against each statistical models that are described in the following subsections.

\subsubsection{Histogram Based Classification Results}
As we see from Table \ref{tab:table2}, OT deviation and DL deviation features performs better than the other features when considered individually. Hence, it would be practical to consider a combination of features and test them against each models. We use different combination of the features and Histrogram is used for the classification. 
\begin{figure}[h]
 \centering
 \includegraphics[width=7.5cm]{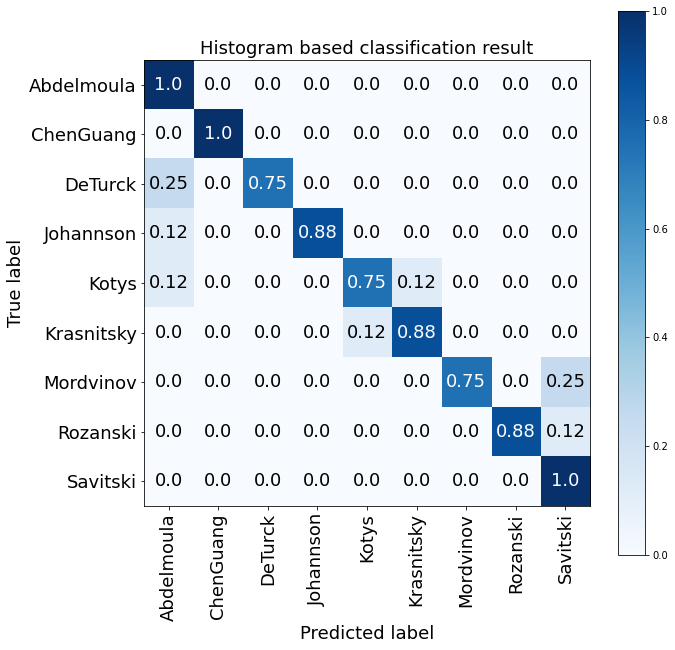}
 \caption{\label{fig:histogram_fusion}Normalized confusion matrix for pianist classification using fusion features for Histogram.}
\end{figure}

\begin{table}[h]
\centering
\resizebox{\columnwidth}{!}{%
\begin{tabular}{|c|c|c|c|}
\hline
\textbf{Feature} & \textbf{Precision} & \textbf{Recall} & \textbf{F-score} \\ \hline
DL+ND            & 0.728              & 0.653           & 0.688            \\ \hline
IOI+DL           & 0.830              & 0.778           & 0.803            \\ \hline
IOI+DL+ND        & \textbf{0.903}     & \textbf{0.875}  & \textbf{0.889}   \\ \hline
OTD+DL+ND        & 0.746              & 0.694           & 0.719            \\ \hline
IOI+OTD+DL+ND    & 0.891              & 0.861           & 0.875            \\ \hline
\end{tabular}%
}
\caption{Pianist identification using Histogram for different combination of features.}
\label{tab:table4}
\end{table}
The best five combination for which the Histogram model produces best result is shown in Table \ref{tab:table4}. The results in bold correspond to the best result. The F1-score is also calculated from the precision and recall. A corresponding confusion matrix for the fusion of IOI deviation, DL deviation and ND deviation is shown in Figure \ref{fig:histogram_fusion}.

\subsubsection{KDE Based Classification Results}
We use Kernel density estimation (KDE) for different fusion features. Table \ref{tab:table2} shows that KDE outperforms the other two distribution models for piano identification considering the individual features. Hence, it is also sensible to experiment with different fusion features to notice if they produce any better result. 
\begin{table}[h]
\centering
\resizebox{\columnwidth}{!}{%
\begin{tabular}{|c|c|c|c|}
\hline
\textbf{Feature} & \textbf{Precision} & \textbf{Recall} & \textbf{F-score} \\ \hline
IOI+DL           & 0.756              & 0.722           & 0.739            \\ \hline
IOI+OTD+DL       & 0.756              & 0.722           & 0.739            \\ \hline
IOI+DL+ND        & \textbf{0.822}     & \textbf{0.792}  & \textbf{0.807}   \\ \hline
OT+OTD+DL+ND     & 0.690              & 0.667           & 0.678            \\ \hline
OT+IOI+OTD+DL+ND & 0.692              & 0.681           & 0.686            \\ \hline
\end{tabular}%
}
\caption{Pianist identification using KDE for different combination of features.}
\label{tab:table5}
\end{table}
We experimented with all the combination of features without any overlap and at least two features are considered. The best five combinations for which the KDE produces the best results are shown in Table \ref{tab:table5}. The IOI deviation feature, fused with dynamic level deviation and ND deviation produces the best result which is shown in bold numbers. 

\subsubsection{GMM Based Classification Results}
We use Gaussian Mixture Model (GMM) based performer identification method using fusion features. We use all the combination of timing deviation features as well as the articulation features.
\begin{table}[]
\centering

\resizebox{\columnwidth}{!}{%
\begin{tabular}{|c|c|c|c|}
\hline
\textbf{Feature} & \textbf{Precision} & \textbf{Recall} & \textbf{F-score} \\ \hline
IOI+DL           & \textbf{0.719}     & \textbf{0.653}  & \textbf{0.684}   \\ \hline
IOI+OTD+DL       & 0.618              & 0.528           & 0.569            \\ \hline
IOI+DL+ND        & 0.618              & 0.528           & 0.569            \\ \hline
OTD+DL+ND        & 0.583              & 0.528           & 0.554            \\ \hline
OT+IOI+OTD+DL+ND & 0.591              & 0.611           & 0.601            \\ \hline
\end{tabular}%
}
\caption{Pianist identification using using GMM for different combination of features.}
\label{tab:table6}
\end{table}

We can see from Table \ref{tab:table6} that, the IOI deviation feature fused with the DL feature outperforms the other fused features. 


\vspace{-5pt}
\section{Analysis and Discussion} \label{discussion}
As discussed in Section 4, there are two main methods used for pianist identification. First, we classify the performers using three distribution models considering the individual features and second, we use a feature fusion method to identify each performers. Table \ref{tab:table2}, shows the pianist identification result for each feature in terms of precision. We observed that KDE performs better for each individual feature except for the OTD deviation feature. No matter which model we use, OT deviation feature tends to perform better out of all other features. The highest precision is achieved using the KDE distribution (0.688). Also in Table \ref{tab:table3}, we show the precision, recall as well as the calculated macro F-score result based on single feature. The result also shows that OT deviation feature outperforms the other features when considered individually with a F-score of 0.677 where the OTD deviation feature performs worst (0.357). To elaborate more, we generate a normalized confusion matrix (Figure \ref{fig:ot_dev_class}) for OT deviation using Kernel density estimation (KDE). The Figure shows that the identification of the performers: Johannson, Kotys and Krasnitsky are the best(0.88) where, the identification of DeTurk, Mordvinov and Savitski are also good (0.75). These performers can be identified correctly most of the time. Rozanski also performs well (0.50) but sometimes confused with Johannson. The identification of Abdelmoula and ChengGuang are mostly unreliable, as we can see their performances are confused with each other as well as other performers. This is due to sharing similar OT deviation feature distributions. Apart from this, the OT deviation feature performs reasonably well for identifying the performers most of the time. However, certain performers can not be identified correctly since they may share the similar feature characteristics.

Considering this problem, we use feature fusion method (Section \ref{feature_fusion_method}) for the performer identification task. Results show that this method works really well identifying the performers correctly. Histogram performs better than other two distribution models and yields a precision of 0.903 for the fusion of IOI, DL and ND deviation features as shown in Table \ref{tab:table4}. Moreover, the confusion matrix in Figure \ref{fig:histogram_fusion} shows that the fusion feature is able to identify each performer more accurately than using only OT deviation feature. Table \ref{tab:table4} also shows that most of the fusion features perform better than the individual features. The features proposed in this study performed better than the features in \cite{masterviolin}, although the identification method as well as some of the features used are quite similar. The probable reason could be that the dataset sizes are different. This could be due to the differences in dataset sizes and different performance features. Moreover, features in this study are extracted from MIDI file whereas they used audio data which might have resulted the deviations to be more noisy. This exhibits the power of our proposed model for the performer classification task.

A potential future direction to improve the performance of the classification would be to consider both low and high level features described in \cite{jsymbolic}. Low level features have received by far the most attention by the MIR researchers whereas the high level features (instruments present, melodic contour, chord frequencies and rhythmic density) have been neglected and mostly used for audio format. Meanwhile, research involving high level features show that they also exhibit a significant discriminating power \cite{McKay2005AutomaticMC}. In addition, a weight optimization can also be done to fuse features with different weights. Another potential interesting future direction could be to segment a sample into parts of equal length and apply the presented methodology to each parts rather than the whole sample. Moreover, features extracted from other data modalities (motion data obtained from sensors) as presented in \cite{beici1,liang2018measurement} could be used to potentially contribute in performer identification task. Furthermore, other classification methods like KNN (K-nearest Neighbors), Decision tree, SVMs and Neural Networks can be tested to improve the classification result. Finally, we aim to use a larger dataset consisting of different composers. 

\vspace{-5pt}
\section{CONCLUSION}
In this paper, we presented an identification method for the recognition of virtuoso pianists using fusion of features extracted from symbolic representation of music. We also constructed a dataset consisting of nine virtuoso pianists, each playing the same piece of music. Our result shows the proposed fusion features to be robust in the exceptionally challenging task of performer identification which normally requires trained/expert human listeners in most situations. The classification accuracy results produced by our model is much higher than the results obtained in previous studies. It has also been demonstrated that the `norm performance' works reasonably well as a reference point. Therefore, it can be sensible to use the norm performance as a reference point where many performances of the same piece are available. The results also show that features related to expressive timing and loudness are the most informative when fused together followed by the aspect of note duration.


\bibliography{icmc2021template}

\end{document}